\documentclass[10pt,letterpaper]{article}
\usepackage{opex3}

\begin{document}

\title{Passive intrinsic-linewidth narrowing of ultraviolet extended-cavity diode laser by weak optical feedback}

\author{Polnop Samutpraphoot,$^{1}$ Sophie Weber,$^{1}$ Qian Lin,$^{1}$ Dorian Gangloff,$^{1}$ Alexei Bylinskii,$^{1}$ Boris Braverman,$^{1}$ Akio Kawasaki,$^{1}$ Christoph Raab,$^{2}$ Wilhelm Kaenders,$^{2}$ and Vladan Vuleti\'{c}$^{1,*}$}
\address{$^1$Department of Physics and Research Laboratory of Electronics,\\ Massachusetts Institute of Technology, Cambridge, Massachusetts 02139, USA}
\address{$^2$TOPTICA Photonics AG, Lochhamer Schlag 19, D-82166 Graefelfing,
Germany}
\email{$^*$vuletic@mit.edu}	
	
\date{\today}

\begin{abstract}
We present a simple method for narrowing the intrinsic Lorentzian linewidth of a commercial ultraviolet grating extended-cavity diode laser (TOPTICA DL Pro) using weak optical feedback from a long external cavity.  We achieve a suppression in frequency noise spectral density of 20 dB measured at frequencies around 1 MHz, corresponding to the narrowing of the intrinsic Lorentzian linewidth from 200 kHz to 2 kHz. The system is suitable for experiments requiring a tunable ultraviolet laser with narrow linewidth and low high-frequency noise, such as precision spectroscopy, optical clocks, and quantum information science experiments.
\end{abstract}




\section{Introduction}
Frequency-stabilized and narrow-linewidth lasers are foundational tools in precision spectroscopy and quantum information science.  They can be used to address narrow transitions in atoms and molecules in order to understand their structure, perform precision measurements of fundamental constants, or manipulate two-level systems as quantum bits of information. Narrow-linewidth ultraviolet (UV) lasers generally require frequency doubling or tripling of a narrow-linewidth infrared system, and such systems are typically expensive and space-inefficient compared to laser diode systems. Traditional extended-cavity diode lasers (ECDLs) \cite{Ricci1995} on the other hand have typical linewidths of a few hundred kHz to a few MHz, with shorter-wavelength systems generally on the broader end \cite{Loh2006}. Numerous schemes exist for narrowing their linewidth.

Optical feedback from a resonator \cite{Dahmani1987} can be used to reduce the linewidth of an ECDL, and recent systems \cite{Labaziewicz2007,Zhao2011,Hayasaka2011,Thompson2012} have demonstrated linewidth narrowing down to 7 kHz. Hertz-level linewidth was achieved with this method using a distributed feedback diode \cite{Zhao2012}. A simpler alternative to extend the optical feedback path length is by means of a fiber and retroreflector, a method used to reduce the linewidth to sub-kHz levels in a 780 nm distributed Bragg reflector laser \cite{Lin2012}, where a 35 dB reduction in high-frequency noise was demonstrated with 3.6 m of optical path length. Concurrently, active linewidth narrowing schemes, such as electronic stabilization to a high-finesse, ultrastable vertical cavity \cite{Ludlow2007,Alnis2008}, or an all-fiber Michelson interferometer \cite{Kefelian2009}, have also achieved hertz-level linewidth. However, due to finite bandwidth for electrical feedback to the diode current, the high-frequency noise is usually not substantially reduced. Such high-frequency laser noise is detrimental to certain applications, such as atom trapping in an optical lattice \cite{Savard1997}, particularly as a source of heating and decoherence \cite{Nazarova2008,Wineland2009}. These schemes are also undesirably complex given the need for high-finesse cavities, with only limited finesse available in the UV.

In this paper, we present a simple, low-cost method for narrowing commercially available ECDLs, with a focus on a UV system. We use a long external cavity formed by a fiber to perform weak optical feedback, as in Ref. \cite{Lin2012}, on a TOPTICA DL Pro laser, and characterize the linewidth over a range of different feedback powers and fiber lengths. In particular, we establish that this secondary optical feedback mechanism is compatible with the primary optical feedback of the ECDL provided by the grating, and we achieve a noise suppression of 20 dB measured at frequencies around 1 MHz. This passive narrowing makes it straightforward to then perform low-bandwidth electrical feedback for further narrowing and stabilization. Indeed, the simple modification to a commercial system presented here is a prototype for a widely affordable kHz-linewidth UV laser system.


\section{Lorentzian Linewidth \& Optical Feedback}

The Lorentzian lineshape of a single-mode laser arises in the frequency domain as the result of random jumps in the phase, or equivalently frequency, of the emitted light. The frequency-noise spectral density $S_{\nu}(f)$ of this random frequency noise is white, and the Lorentzian model is valid in a spectral region for which the amplitude of the frequency noise is smaller than the spectral frequency \cite{DiDomenico2010}. We can then obtain the Lorentzian linewidth of the laser $\Gamma_{L}$ with the simple relation

\begin{equation}\label{eq1}
\Gamma_{\rm L}=\pi S_{\nu 0}
\end{equation}

\noindent where $S_{\nu 0}$ is the white frequency noise level in the relevant spectral region, measured in ${\rm Hz}^2/{\rm Hz}$ \cite{DiDomenico2010}. 

For the system presented in this paper, the Lorentzian model would be valid for frequencies greater than $\sim 200$ kHz, given by the frequency noise amplitude of the ECDL. However, $S_{\nu}(f)$ is dominated by other sources of noise up to $\sim 1$ MHz. As a result, the Lorentzian contribution to $S_{\nu}(f)$ is taken to be for frequencies greater than $\sim 1$ MHz. It is relatively easy to suppress noise below $\sim 1$ MHz using servo-electronics with appropriate bandwidth. For higher frequency noise, it is necessary to rely on alternative noise suppression methods.

One can apply weak optical feedback to suppress this high-frequency noise and narrow the Lorentzian linewidth using a long external cavity, whose long optical path length results in constructive interference for light with long coherence length, corresponding to a narrower linewidth.  Although we expect a longer cavity to perform better in narrowing, the maximum length is limited due to a noise peak at the free spectral range (FSR) of the long external cavity, as shown in the following section of this paper. Indeed, for a sufficiently long cavity with a correspondingly long light round-trip time, and thus a short FSR, there will be an increase in noise in the frequency range where we desire low-noise operation of the laser.

The behavior of the laser also depends on feedback power and phase. 
According to Ref. \cite{Tkach1986}, the behavior can be classified into five regimes depending on the ratio $p$ of feedback to overall output power. Of particular interest in this paper is the regime for which linewidth narrowing is expected to occur, at $p< -39$ dB.
When $-45$ dB $<p< -39$ dB, the system exhibits sensitivity to phase, i.e, the feedback phase from the long external cavity needs to be tuned for constructive interference with
the grating cavity feedback phase to produce narrowing.  The sensitivity increases with feedback power, and we expect the frequency noise suppression to increase with $p$ while $p<-39$ dB, beyond which the laser becomes multimode.  In this experiment, we operate in this regime, up to the edge of instability.

\section{Measurements of Linewidth}

\begin{figure}[htb]
\centering\includegraphics[width=9.2cm]{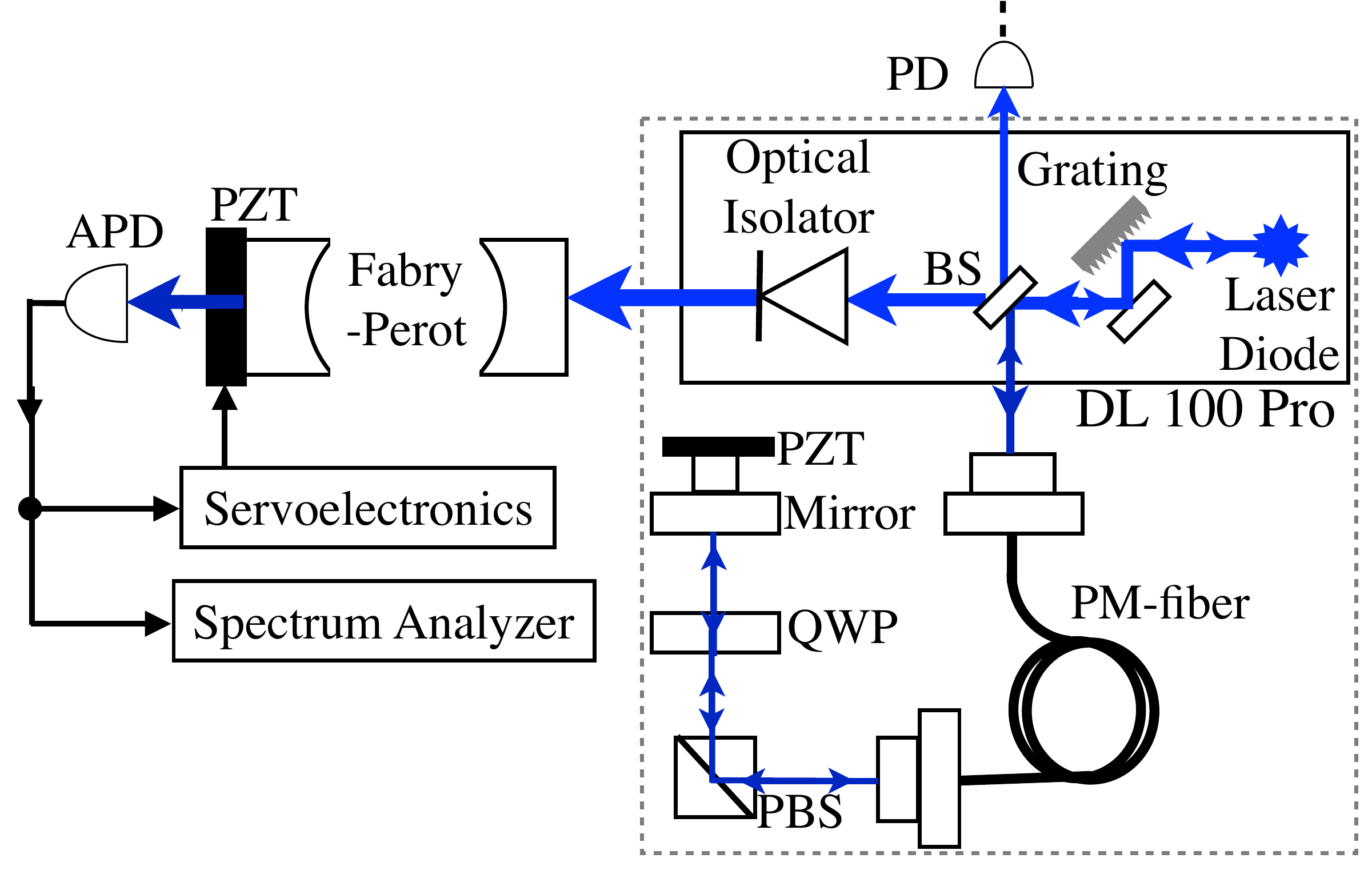}
\caption{Schematic diagram of the apparatus. The complete narrow linewidth system, consisting of the TOPTICA DL Pro supplemented by the long external cavity, is inside the dotted-line box, and the measurement chain is outside. The following abbreviations were used: PM for polarization-maintaining, PBS for polarizing beamsplitter, QWP for quarter-wave plate, PZT for piezoelectric transducer, and APD for avalanche photodiode. 
\label{fig1}}
\end{figure}

The experimental apparatus is shown in fig.~\ref{fig1}.
The laser is a 411 nm TOPTICA DL Pro with a built-in grating in Littrow configuration used for line narrowing and wavelength tuning \cite{Ricci1995}.  

The feedback from the long external cavity is obtained using a beam splitter (BS) which picks off 10$\%$  of the ECDL's light into a single-mode polarization maintaining (PM) fiber with coupling efficiency of $\sim$ 40$\%$. A mirror is placed after the output coupler to reflect the feedback light back into the fiber and laser. We control the feedback power with a quarter-waveplate (QWP) and a polarizing beam splitter cube (PBS) placed between the fiber output coupler and the feedback mirror. The feedback power is monitored using a photodiode (PD) as shown in fig.~\ref{fig1}. With this setup, the ratio of feedback power to power measured after the optical isolator (71\% transmission) reaches a maximum of -29 dB. 

The external-cavity length is easily varied using fibers of different lengths. In this experiment, we alternate between a 1 m fiber and a 4 m fiber, with total round-trip optical path lengths including free-space distances, of 4.6 m and 13.6 m respectively, corresponding to FSRs of the long external cavity of 65 MHz and 22 MHz respectively. The path length can be fine-tuned on the order of a wavelength by changing the position of the feedback mirror with a piezoelectric transducer (PZT), thus tuning the laser frequency.  

In order to characterize the laser frequency noise, we convert it to intensity noise by passing laser light through a Fabry-Perot cavity (FP) at half-transmission.  The intensity noise is then detected on a photodiode. The length of the FP is stabilized with respect to the laser frequency using a side-of-fringe locking technique. For a locking point in the linear region of the FP's transmission spectrum, the measured intensity signal is directly proportional to frequency noise. The bandwidth of the servo loop is set to $\sim$ 1 kHz to ensure that the high-frequency components that are of principal interest here pass undistorted by the active feedback loop.

The FP side-of-fringe linear conversion factor $g$ in ${\rm V}/{\rm Hz}$ from frequency fluctuations to measured intensity fluctuations, converted by the photodiode into an output voltage, is obtained by measuring the full-width at half-maximum (FWHM) of the FP as the laser is scanned across resonance.  In this experiment, we use two FP cavities, with FWHM of 40 MHz and 7.5 MHz respectively.

The Lorentzian linewidth can be determined from the white part of the noise spectrum $S_{\nu 0}$ (eq.~\ref{eq1}), converted from a correspondingly white power spectrum $S_{P0}$ (measured in V$^{2}$/Hz) using the calibrated constant of proportionality $g$ described above:

\begin{equation}\label{eq2}
S_{\nu 0} = \frac{S_{P0}}{g^{2}}
\end{equation}

We observe the onset of the white-noise region $S_{P0}$ at $\sim$ 100 kHz and take the level between 500 kHz and 2 MHz, well below the roll-off of the narrow FP and the APD, for the calculation.
The Lorentzian linewidth of the TOPTICA DL Pro laser without feedback is determined from this method to be $\sim 170$ kHz.  

The absolute value of the bare laser linewidth obtained from the aforementioned calculation relies on the spectral flatness of the measurement response; indeed, the conversion factor $g$ is measured near DC, while the frequency noise spectrum is measured in the MHz range.  As a cross-check, we perform a high-frequency calibration of $g$. We frequency modulate the diode current at $f_{m}$ ($>$ FWHM of the FP), with a modulation depth $\beta$ as measured from the relative heights of the laser carrier and modulated sidebands observed in transmission through a FP. The amplitude of frequency modulation is then simply $\Delta f=\beta f_{m}$. At half-transmission through the FP, the frequency modulation produces an intensity signal $g\Delta f$, whose measured power density is thus $S_{P}(f_{m})=(g\Delta f)^{2}/f_{RBW}$, where $f_{RBW}$ is the chosen resolution bandwidth of the spectrum analyzer. We can thus obtain a value of $g$ measured at relevant frequencies, now yielding a measurement of the bare laser linewidth of $245$ kHz.

Taking the intrinsic Lorentzian linewidth of the ECDL as the average of the two values determined above, we obtain $210 \pm 50$ kHz, consistent with \cite{Zhao2010}. Absolute values of linewidth calculated henceforth are taken relative to this level.  

\begin{figure}[htb]
\centering\includegraphics[width=13cm]{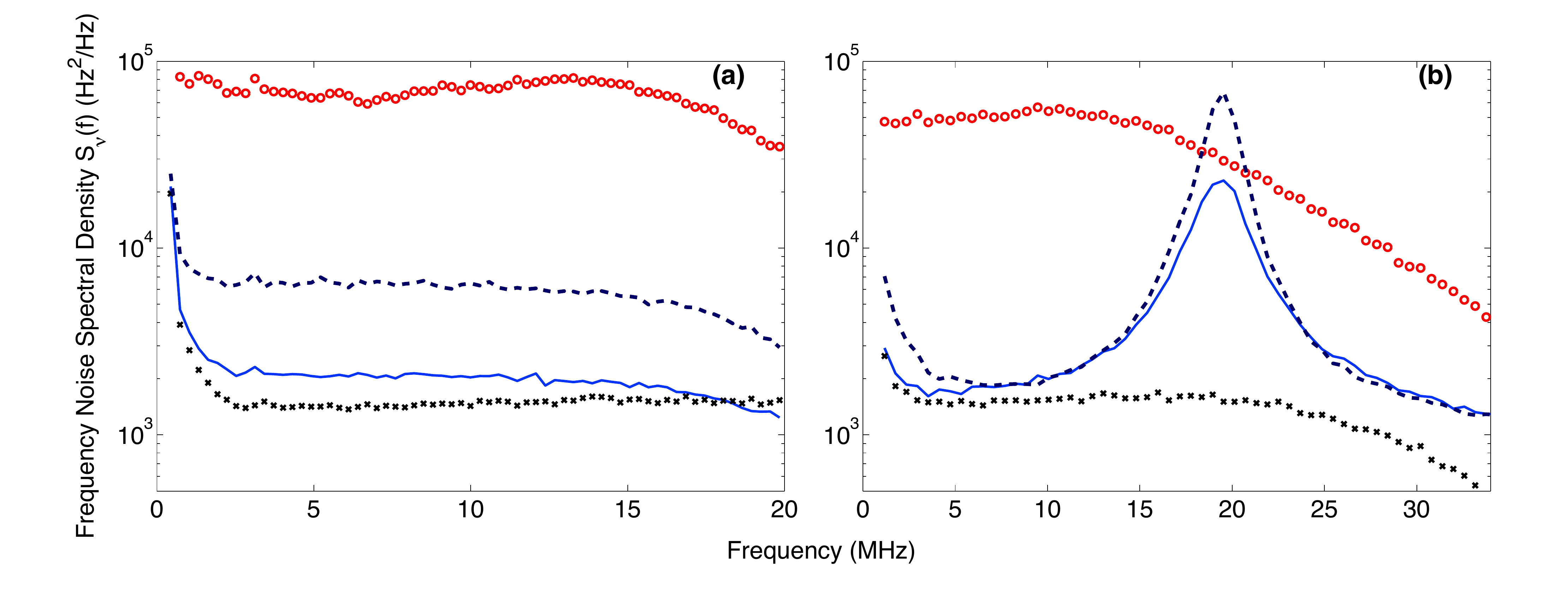}
\caption{Frequency noise spectral density without feedback (red circles), with intermediate feedback level (dark blue dashed), minimum frequency noise corresponding to highest stable feedback level (light blue solid), and amplitude noise baseline (black crosses) for 1m fiber (a) and 4m fiber (b). Feedback values for intermediate and minimum frequency noise are -44 dB and -40 dB for the 1 m fiber (a), and -43 dB and -39 dB for the 4 m fiber (b). The FP used for frequency analysis here has FWHM = 37 MHz, causing all but the amplitude noise traces to roll-off around 18 MHz. Noise peaks in panel (b) are at the FSR of the long external cavity.
\label{fig2}}
\end{figure}

Typical data for the frequency noise density $S_{\nu}(f)$ are shown in fig.~\ref{fig2}. These data are taken with the broader FP (FWHM = 37 MHz) so as to measure high-frequency features, though linewidths quoted in this paper are from data taken with the narrower FP (FWHM = 7.5 MHz), for better accuracy. The frequency noise suppression is shown to be at best 20 dB for both a 1 m fiber and a 4 m fiber, corresponding to Lorentzian linewidths of 2 kHz. In both cases, the minimum linewidth achieved is for a feedback ratio of $p=-39$dB, where a stronger feedback ratio produces multimode behaviour. To ensure that the measurement is not dominated by direct intensity fluctuations of the laser light, we characterize this amplitude noise by measuring the power spectral density of light directly incident on the APD (black crosses in fig.~\ref{fig2}).

The broader FP gives sufficient bandwidth to observe noise features at the FSR of the cavity formed by the 4 m fiber (Fig.~\ref{fig2}b).  We find a 10MHz-wide peak in frequency noise at 20MHz, which is the inverse round-trip time of light in the external cavity.  This noise peaking effect may limit the usefulness of the narrowing in applications that require low-noise operations at frequencies on the order of the FSR of the long external cavity. This additional noise is reduced with higher feedback, as demonstrated by the lower peak value in fig.~\ref{fig2}b. Nonetheless, the maximum value of this noise at all feedback levels is at most 30\% above the white noise level corresponding to the intrinsic linewidth of the laser.

Note also that in the traces corresponding to a narrowed linewidth, the sloped features below $\sim$ 1 MHz reveal part of the frequency noise associated with the non-Lorentzian contribution to the linewidth.  Although we neglect it in our linewidth calculation, this low-frequency noise is also partially reduced with increased optical feedback. It can also be easily suppressed via electronic feedback when the laser is locked to a stable reference such as an optical cavity or an atomic line.

Fig.~\ref{fig2} also illustrates that the measurement of narrowed linewidth may be limited by the laser amplitude noise (black traces in fig.~\ref{fig2}).  We thus increase the gain $g$ with a narrower FP, but at the expense of reducing the overall bandwidth of the measurement, limited to the FWHM of the FP.  We systematically characterize the noise suppression by the external cavity for different amounts of feedback power and establish a narrowing limit within our measurement capabilities. Lorentzian linewidths are calculated using the white region of the spectra $S_{\nu 0}$ (not shown), as inferred from a fit of the roll-off of the measurement FP between 0.8MHz and 7MHz.

\begin{figure}[h!]
\centering\includegraphics[width=12cm]{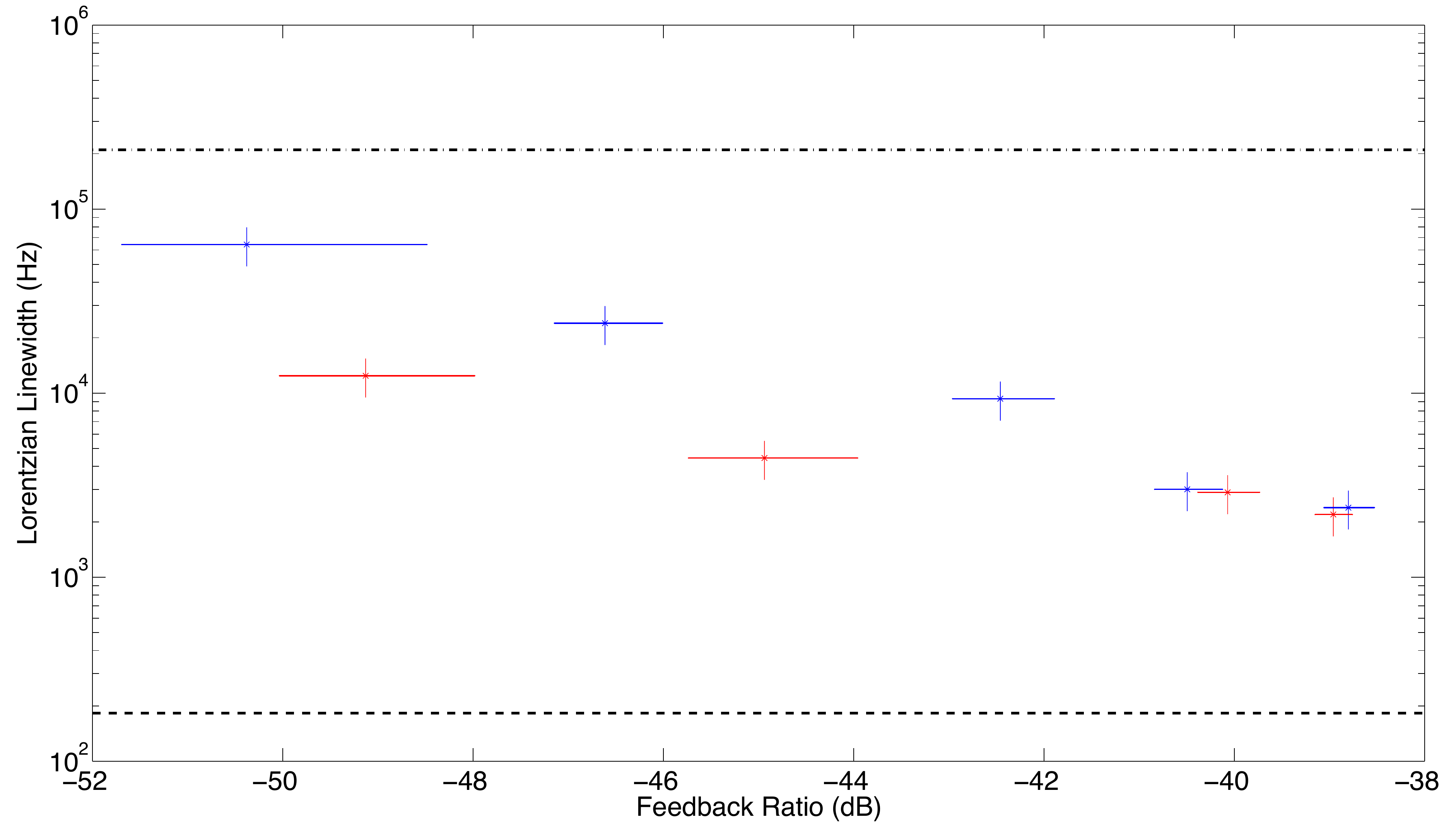}
\caption{Lorentzian linewidth as a function of feedback power ratio $p$ from long external cavity using a 1 m fiber (blue) and a 4 m fiber (red). The linewidth narrowing is limited in both cases by multimode instability as $p$ approaches the critical value of $-39$ dB. The top dash-dotted line represents the ECDL linewidth of 210 kHz without feedback from the long cavity, while the bottom dashed line represents the resolution limit due to amplitude noise. The FP used here has FWHM = 7.5 MHz.
\label{fig3}}
\end{figure}

Fig.~\ref{fig3} shows the Lorentzian linewidth as a function of the feedback power ratio, for both fiber lengths.
The narrowing capability is confirmed to be better for a longer cavity, in agreement with theoretical expectations \cite{Henry1986} and previous measurements on DBR lasers \cite{Lin2012}. At low and intermediate feedback levels, we observe that for equal feedback ratios the linewidths follow the expected square law relation with feedback cavity length, wherein a ratio in the cavity length of $\sim 3$ leads to a linewidth lower by almost an order of magnitude. At feedback levels close to the instability ($p\approx$ -39 dB), the longer external cavity appears to saturate at noise levels equivalent to that of the shorter external cavity, perhaps due to noise level with multimode behaviour. 

For comparison, and as a demonstration of the broad applicability of this narrowing method to a range of ECDL systems, we reproduced similar results in a nearly identical setup with a 1156 nm TOPTICA DL Pro laser. There we observed up to a 16 dB reduction in frequency noise, measured in the range 100 kHz to 1 MHz, using an external cavity with a 10 m optical path length built from a 3 m fiber.

\section{Feedback Ratio, Phase and Stability}

As demonstrated above, the feedback ratio $p$ plays an important role in the stability of the laser diode with optical feedback, with feedback levels $p>\sim -39$ dB producing a sharp transition to the unstable regime.  In our system, as $p$ approaches -39 dB from below, the laser becomes highly sensitive to the feedback phase, which coincides with Regime II of Ref. \cite{Tkach1986} ($p<-45$ dB in \cite{Tkach1986}), while the features associated with Regime III ($-45$ dB $<p< -39$ dB in \cite{Tkach1986}), such as the ability to operate single-mode with narrow linewidth and insensitivity to phase, are not apparent in this ECDL system. If the feedback power is increased further, the laser enters Regime IV ($p>-39$ dB both in \cite{Tkach1986} and in our system), exhibiting coherence collapse, which cannot be remedied by tuning the feedback phase. The linewidth narrowing is limited in the current system by this instability.

There are two external frequency-tuning elements in this modified ECDL: the grating and the long external cavity. Single-mode operation requires that optical feedback from the long external cavity match the phase of the optical feedback from the grating. Indeed, tuning the relative position of the grating and feedback mirror for constructive interference was necessary before obtaining each spectral data set. This matching of both cavities for constructive interference was affected by the passive stability of the relative lengths of the two cavities; typically a few minutes in this setup. Additionally, this constraint on the relative phases of optical feedback implies that locking the laser to an external reference requires active stabilization of both frequency-tuning elements simultaneously.

The passive stability of the ECDL, here defined as the time between mode-hops on the order of the FSR of the long external cavity, is limited in great part by the relative drift of the optical path lengths of the two feedback paths. The dominant contribution was found to be changes in the index of refraction of air resulting from environmental pressure changes.  A drift of the absolute laser frequency of 100 MHz/hour can be measured in the TOPTICA DL pro system, in the absence of feedback from the long fiber cavity, consistent with environmental pressure drifts on the order of 0.5 mbar/hour. Given the FSR of the long external cavity of a few 10's of MHz and the requirement of constructive interference between the two frequency tuning elements, it is not surprising that the passive stability of the system with long cavity feedback be limited to a few minutes. This may naturally be remedied by active stabilization of the relative length of the two feedback cavities, or by stabilizing the pressure under which the ECDL operates. As an example of the former, one untested proposal is to dither the grating cavity length so as to modulate the laser frequency on the order of the FSR of the external cavity. The resulting modulation in feedback phase could produce an intensity modulation signal large enough to lock the grating cavity length relative to the fiber cavity length. 
  
\section{Conclusion}
We have implemented weak optical feedback using a long external fiber cavity on an ultraviolet grating ECDL.  We have achieved a 20 dB reduction in high-frequency noise with long external cavities, corresponding to narrowing the Lorentzian contribution to the linewidth from 210 kHz to 2 kHz.  Additional noise suppression is limited by instability of the laser at high feedback levels. 
Due to its capability for substantial narrowing of the intrinsic laser linewidth of an ECDL, this simple design, consisting of a straightforward modification to a commercially available system, is promising for experiments requiring high precision at costs significantly lower than other currently available solutions. This paper also supplements recent results on the passive narrowing of a DBR laser using the same technique, and demonstrates its broad applicability to achieving low-linewidth laser diode systems.

\section{Acknowledgements}

This work was supported in part by the NSF, the NSF sponsored Center for Ultracold Atoms, the AFOSR, and DARPA. P.S., S.W., and Q.L. acknowledge support from the Undergraduate Research Opportunity Program at MIT, D.G., A.B. and B.B. from the National Science and Engineering Research Council of Canada. 

\end{document}